\documentclass[prb,preprint,showpacs]{revtex4}

\usepackage{graphicx}

\begin{document}

\widetext

\title{Dynamical model of the dielectric screening of conjugated polymers}

\author{William Barford$^{1*}$, Robert J. Bursill$^2$ and
David Yaron$^{3}$}

\affiliation{ $^1$Department of Physics and Astronomy, University
of Sheffield, Sheffield, S3 7RH, United Kingdom.
\\
$^2$School of Physics, University of New South Wales, Sydney, NSW
2052, Australia.
\\
$^3$Department of Chemistry, Carnegie Mellon University, Pittsburgh, PA 15213, U.S.A. }

\begin{abstract}

A dynamical model of the dielectric screening of conjugated
polymers is introduced and solved using the density matrix
renormalization group method. The model consists of a line of
quantized dipoles interacting with a polymer chain. The polymer is
modelled by the Pariser-Parr-Pople (P-P-P) model. It is found
that: (1) Compared to isolated, unscreened single chains, the
screened $1^1B_u^-$ exciton binding energy is typically reduced by
ca.\ 1 eV to just over 1 eV. (2) Covalent (magnon and bi-magnon)
states are very weakly screened compared to ionic (exciton)
states. (3) Screening of the $1^1B_u^-$ exciton is closer to the
dispersion than solvation limit.

\end{abstract}


\maketitle

\section{Introduction}

The intra-chain excitations of a conjugated polymer in the solid
state are often strongly screened by the environment of the other
polymers. This screening causes significant reductions of the
ionic single-chain excitation energies and charge gaps, and thus
single-chain calculations are expected to considerably
over-estimate the actual excitation energies of polymers in the solid
state. In particular, the energy of a point charge is typically
reduced by approximately $1$ eV, while the excitation energy of
the lowest dipole allowed exciton is reduced by approximately
$0.5$ eV, implying a deviation of approximately $1.5$ eV in the
single-chain calculated exciton binding energy in comparison to
 the solid state measurements.\cite{yaron0,yaron}

It is useful to consider two limiting cases for the interaction
between the electron-hole pair and the polarization of the
dielectric medium, namely solvation-like versus dispersion-like
interactions. By solvation, we mean a polarization that develops
around the electron and hole and follows the motion of these
charged species {(that is, the dielectric is fast compared
to the electron-hole motion)}. In this case, the effects of the
polarization can be absorbed into a {static} screened
electron-hole Coulomb interaction. By dispersion, we mean the
interaction between fluctuating dipoles, such as may arise between
the fluctuating dipole of the bound electron-hole pair in an
exciton and the dipoles of the dielectric medium. The interactions
become dispersion-like when the time scale of the electron-hole
motion is comparable to or less than the time scale of the
dielectric response {(that is, the electron-hole motion is
fast compared to the dielectric)}. In this case, the dielectric
polarization does not identically follow the electron-hole motion,
but rather fluctuates in a manner that is correlated with the
motion of the electron-hole pair. {Now, the effects of the
polarization may be modelled by a dynamically screened
electron-hole Coulomb interaction.}

The nature of the screening thus depends on the relative time
scale of the electron-hole motion, which is approximately
inversely proportional to the exciton binding
energy\cite{footnote3}, versus the time scale of the dielectric
polarization, which is inversely proportional to the optical gap
of the dielectric medium. In inorganic semiconductors, the exciton
binding energy is one to two orders of magnitude smaller than the
optical gap. The polarization is then much faster than the
electron-hole motion, such that it can follow the electron-hole
motion and effectively screen the Coulomb interaction.\cite{knox}
In organic semiconductors, the exciton binding energy is
comparable to the optical gap and we expect the interaction
between the exciton and the dielectric polarization to have a
dispersion-like character. In this regime, there is no {\em a
priori} reason to suppose that a static dielectric response
function will correctly describe the relevant physics\cite{yaron}.

In this paper we present a dynamic model of the dielectric -
making no assumptions about time scales - with which we attempt to
understand the screening of the key excited states of conjugated
polymers.  We consider the r\^ole of lengths scales (namely, the
electron-hole separation versus the length scales in the
monopole-dipole interaction), time scales (namely, exciton
excitation energy versus binding energy), and ionicity versus
covalency in the excited states.  The correlated electron dynamics
within a single chain are modelled by the Pariser-Parr-Pople
(P-P-P) model, while the dielectric is treated as a linear chain
of quantized dipoles.  This model is solved essentially exactly
via the density matrix renormalization group (DMRG)
method.\cite{white}

The plan of this paper is as follows: in the next section the model is
introduced and parameterized.  In the following section we discuss our
results, and finally conclude.

\section{Dynamic dielectric model}

The model is designed to represent the dielectric response of an
array of conjugated polymers parallel to  the solute (or test)
chain. The dielectric function, $\epsilon(\omega)$,  of an
assembly of polymers is\cite{ziman},
\begin{equation}\label{}
    \epsilon(\omega) = 1 + \frac{N q^2}{\epsilon_0 m} \sum_n
    \frac{f_n}{\omega_n^2 - \omega^2},
\end{equation}
where $\omega_n$ and $f_n$ are the frequency and oscillator
strength for the $n$th transition, $q$ and $m$ are the electronic
charge and mass, $\epsilon_0$ is the dielectric constant of
free-space, and $N$ is the number of polymers per unit volume. In
this paper we consider a `two-state' model, namely we only
consider excitations of the dielectric from the ground state to
the first dipole-allowed state (e.g. the $1^1B_u$ state in
centro-symmetric polymers). Then, using the oscillator sum rule,
\begin{equation}\label{}
    \sum_n f_n = 1,
\end{equation}
where,
\begin{equation}\label{}
    f_n = \frac{2 m \omega_n \langle GS |\hat{\mu} |n \rangle^2 }{q^2
    \hbar}
\end{equation}
we have
\begin{equation}\label{Eq:4}
    \epsilon(\omega) = 1 + \frac{N}{\epsilon_0}
    \frac{2 \langle\hat{\mu} \rangle^2 \omega_0}{\hbar(\omega_0^2
- \omega^2)}.
\end{equation}
In this expression, $\langle \hat{\mu} \rangle$ and $\omega_0$ are the
transition dipole moment and frequency of the excited state.
$\hbar\omega_0$ corresponds to the optical gap of the polymers. We
note that for $\omega >> \omega_0$ Eq.\ (\ref{Eq:4}) becomes,
\begin{equation}\label{}
    \epsilon(\omega) = 1 -  \frac{\omega_p^2}{\omega^2},
\end{equation}
where $\omega_p^2 = N q^2/\epsilon_0 m$ is the plasma frequency.
This frequency is typically $10-20$ eVs - much larger than the
optical gap or the energy scale of the electron-hole motion - so
this limit need not concern us here.

Since the dipole moment of the chain is parallel to the chain
axis, we consider a line of longitudinal dipoles, each oscillating
with a characteristic frequency, $\omega_0$.\cite{footnote6}  The
dipole at $x_j$ interacts with a point-charge on the solute chain
at $x_i$ via the monopole-dipole interaction,
\begin{equation}\label{}
  U_s(x, r) =  \frac {q x \mu}{4\pi\epsilon_0 (x^2 + r^2)^{3/2}},
\end{equation}
where $\mu$ is the dipole moment, $x = (x_j -x_i)$ and $r$ is the
normal distance of the line of dipoles  from the chain, as
illustrated in Fig.\ 1.

\begin{figure}[tb]\label{Fig1}
\begin{center}
\includegraphics[scale = 0.8]{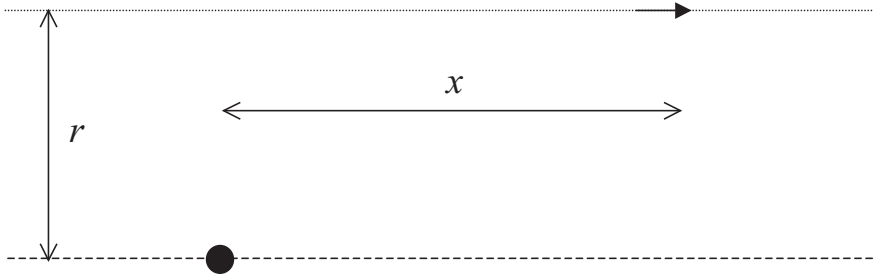}
\end{center}
\caption{The polymer chain (dashed line), with a monopole at $x_i$
(circle) and the line of longitudinal  dipoles (dotted line), with
a dipole at $x_j$ (arrow).}
\end{figure}

A realistic model of the dielectric would consist of a
three-dimensional array of lines of dipoles  surrounding the
solute chain.  However, such  a model would not be easily soluble by the
DMRG method.  Instead, we consider a one-dimensional model and
derive an effective interaction between the monopoles and the
three-dimensional array of dipoles by integrating over a cylinder
with inner and outer radii of $r_1$ and $r_2$, respectively:
\begin{eqnarray}\label{Eq:7}
    U(x;r_1,r_2) = &&\int_{r_1}^{r_2} \rho 2 \pi a r  U_s(x,r)\textrm{d}r \\ \nonumber
  =  &&\left( \frac{q}{4\pi\epsilon_0}\right)\left(\frac{2\pi a}{r_1^2}\right) \left(\rho r_1^2 \mu \right)
\left[\sin\left(\tan^{-1}\left(\frac{x}{r_1}\right)\right)-\sin\left(\tan^{-1}\left(\frac{x}{r_2}\right)\right)\right].
\end{eqnarray}
Here, $\rho$ is the number density of dipoles and $a$ is the
linear separation (for convenience taken to  be the separation
between the sites on the chain). The purpose of this integration
procedure is that it performs a type of average over the distances
between the line dipoles and the solute chain.  However, $r_1$ is
still the effective length scale in the interaction, as shown in
~Fig.\ 2, which shows $U(x;r_1,r_2)$ for different $r_2$ for fixed
$r_1$. For both $r_2 = 8$ $\AA$ and $16$ $\AA$ $U(x;r_1,r_2)$ is
strongly peaked at approximately the value of $r_1$. Thus, using a
single line dipole a distance $r_1$ away would give very similar
results to the cylinder of inner radius $r_1$.

\begin{figure}[tb]\label{Fig2}
\begin{center}
\includegraphics[scale = 0.8]{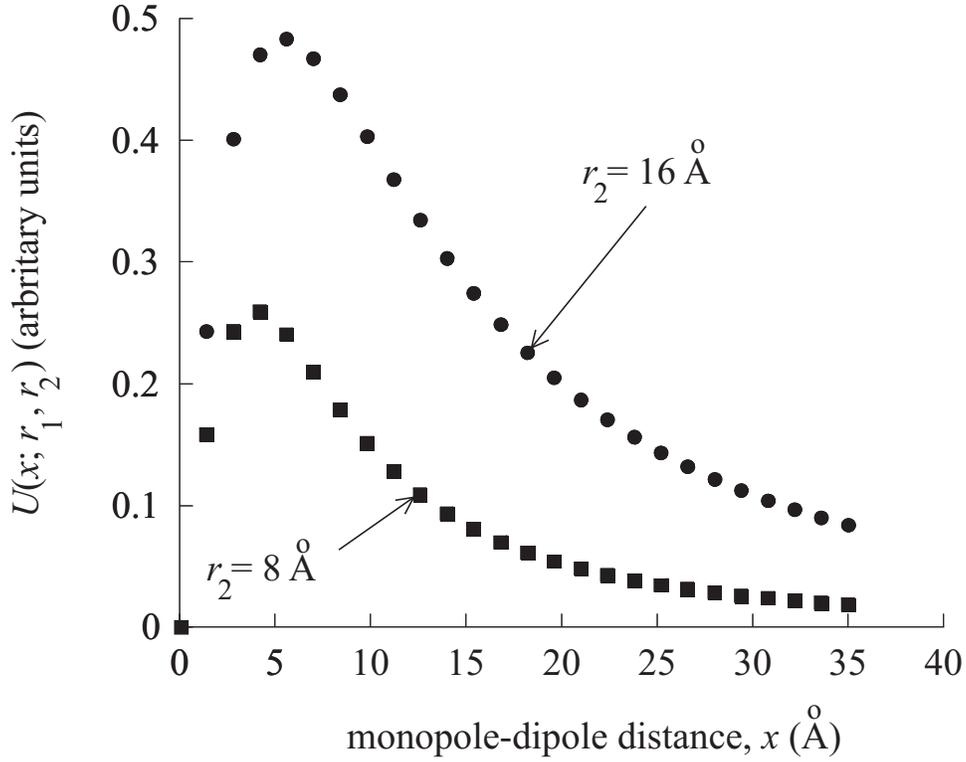}
\end{center}
\caption{The integrated monopole-dipole interaction,
$U(x;r_1,r_2)$, for $r_1 = 4$ $\AA$, with $r_2 = 8$ $\AA$
(squares) and $r_2 = 16$ $\AA$ (circles).}
\end{figure}

The Hamiltonian is thus,
\begin{equation}\label{eq3}
    H = H_e + H_d + H_{e-d},
\end{equation}
where,
\begin{eqnarray}
H_e   = -  \sum_{i \sigma} t_i(c_{i\sigma}^{\dagger} c_{i+1\sigma} +
c_{i+1\sigma}^{\dagger} c_{i \sigma})
 + U \sum_i \left(\hat{N}_{i\uparrow}- \frac{1}{2}\right) \left( \hat{N}_{i\downarrow}- \frac{1}{2}\right) +
\frac{1}{2} \sum_{i \ne j} V_{ij} (\hat{N}_i-1)(\hat{N}_{j}-1),
\end{eqnarray}
is the P-P-P model, describing the interacting electronic degrees of freedom.
$ c_{i \sigma}^{\dagger}$ creates an electron with spin $\sigma$
in the $\pi$-orbital on site $i$ and $\hat{N}_i = \sum_{\sigma} c_{i
\sigma}^{\dagger} c_{i \sigma}$. $V_{ij}$ is the Ohno-Coulomb repulsion,
\begin{equation}\label{eqn5}
V_{ij} = \frac{U}{ \sqrt{ 1 +  \epsilon \beta r_{ij}^2 }},
\end{equation}
where $\epsilon$ is the dielectric constant (generally taken to be $1$), the bond lengths
are in $\AA$ and $\beta = (U/14.397)^2$. The  bond
lengths, $a$, used in the evaluation of $V_{ij}$ are $1.4$ $\AA$, and the bond angle is $180^0$.
The
transfer integral is $t_i = t(1 + (-1)^i\delta)$, where
$\delta$ is the bond dimerization parameter.

The Hamiltonian,
\begin{equation}\label{}
 H_d = \sum_j \hbar \omega_0 b^{\dagger}_j b_j,
\end{equation}
describes the oscillating dipoles. $b^{\dagger}_j$ is a
boson operator creating a quantum  of energy $\hbar \omega_0$ in a
linear harmonic oscillator located at $x_j$.  This term describes the non-interacting self-energy
of the oscillating dipoles.  However, in this model we neglect the dipole-dipole interactions, as
in one dimension this induces an unphysical  spontaneous dipole moment.  Thus, the classical limit
(defined as $\omega_0 \rightarrow 0$) is not completely described by this model.

Using Eq.\ (\ref{Eq:7}), the monopole-dipole Hamiltonian is,
\begin{equation}\label{Eq 7}
    H_{e-d} = \sum_{ij}\left( \frac{q^2}{4\pi\epsilon_0}\right)\left(\frac{2\pi a}{r_1^2}\right)
\left[\sin\left(\tan^{-1}\left(\frac{x}{r_1}\right)\right)-\sin\left(\tan^{-1}\left(\frac{x}{r_2}\right)\right)\right]
(\hat{N}_i-1) \lambda \frac{(b^{\dagger}_j + b_j)}{2},
\end{equation}
where we define the dimensionless monopole-dipole coupling constant, $\lambda$, as,
\begin{equation}\label{eq:8}
  \lambda =  \rho r_1^2 \langle \hat{d}\rangle
\end{equation}
and we have used
\begin{equation}\label{}
    \hat{\mu} = q\langle \hat{d}\rangle \frac{(b^{\dagger}_j +
    b_j)}{2}.
\end{equation}
$q\langle \hat{d}\rangle$ is the transition dipole moment of the solvent
chains associated with the $1^1B_u^-$ exciton\cite{footnote4} and $(b^{\dagger}_j + b_j)/2$ is  the
dimensionless operator
corresponding to the displacement of the oscillator.

In the absence of electron-electron interactions and for only
local monopole-dipole interactions the  Hamiltonian, $H$, is the
Holstein model, which is widely used to study charge-density-wave
phenomena in molecular solids.  The monopole-dipole
interactions imply that $H$ possesses a special type of
particle-hole symmetry.  In particular, $H$ is invariant under the
simultaneous particle-hole transformation of\cite{footnote1}
\begin{equation}\label{}
       c_{i\sigma}^{\dagger} \rightarrow (-1)^i c_{i\bar{\sigma}}
\end{equation}
and a reversal of parity in the boson operator,
\begin{equation}\label{}
    b^{\dagger}_j \rightarrow -b^{\dagger}_j.
\end{equation}

$\hbar \omega_0$ is the excitation energy of the $1^1B_u$ exciton
and $q\langle \hat{d}\rangle$ is its  corresponding transition
dipole moment. These are determined by solving the single chain
P-P-P model in the absence of monopole-dipole interactions. $\rho$
is the only adjustable parameter in the model, and is adjusted to
reproduce the solvation energy of a point charge found by Yaron
and Moore \cite{yaron} for particular values of the P-P-P
parameters, namely, $t$, $U$ and $\delta$. Having fixed $\rho$ for
one set of P-P-P parameters  the model can be transferred to
another set of P-P-P parameters, by recalculating the unscreened values of $\hbar\omega_0$
and $\langle \hat{d}\rangle$, and the solvation of states as a
function of these parameters can be monitored.

\subsection{Parametrization of the model}

Moore and Yaron\cite{yaron} studied solvation effects in polyacetylene by
surrounding a central solute chain by increasing numbers of solvent chains,
arranged as in the crystal structure of ref\cite{fincher}, and extrapolating to an
infinite system. The central chain was treated with a P-P-P model with
parameters, $t=2.4045$ eV, $\delta=0.0734$, $U=11.13$ eV (corresponding to
$t_d = 2.581$ eV and $t_s = 2.228$ eV) and the solvent chains were treated
in an independent-electron approximation. Coulomb interactions were
included between all chains. Replacing the central solute chain with a
point charge led to an estimate of about 1 eV for the solvation energy of a
point charge. In the assumption of an infinitely fast dielectric, where the
solvent polarization is equilibrated to the instantaneous position of the
electron and hole, the solvation energy of a well separated electron and
hole is about 1.9 eV. Using a model that approximates the dynamic response
of the dielectric, the free electron and hole become dressed by the
dielectric response of the solvent chains to form polarons and the
solvation energy drops to about 1.5 eV.
Using the {\em ab initio} DFT-GWA-BSE method, Bobbert and co-workers calculated a
point charge screening in polythiophene of $0.55$ eV, i.e. $1.1$
for the charge gap\cite{bobbert}.

Notice that for the parameter values used by Moore and Yaron
the P-P-P model is strongly correlated\cite{karp2}. The triplet, $1^3B_u^+$,
state is more properly described as a magnon than the $n=1$
Mott-Wannier triplet exciton. Similarly, the $2^1A_g^+$ state is
not the $n=2$ Mott-Wannier singlet exciton, but a bi-magnon, whose
energy lies below the $n=1$ singlet exciton, namely the $1^1B_u^-$ state.\cite{barford02a, barford02b}
The unscreened energies of a $50$ site chain are shown in Table I.

We solve Eq. (\ref{eq3}) by the DMRG method\cite{white} using a
program well-tested on similar electron-boson
models\cite{barford_phonon}. Since the boson energy scale, $\hbar
\omega_0$, is comparatively large compared to the electronic
excitation energies, we found that only one boson per site is
required for the excitation energies to converge to within $0.01$
eV. The number of superblock states was typically $\sim 100,000$
and one finite-lattice sweep at the target chain size was always
performed to ensure convergence.

Taking $r_1 = 4$ $\AA$, $r_2 = 16$ $\AA$ and $\lambda = 0.14$, we obtain a
screening energy of $0.81$ eV for the point charge.  This is intermediate
between the values of Bobbert\cite{bobbert} and Moore-Yaron\cite{yaron} and
is close to the charged-polaron result of $0.88$ for $50$
sites.\cite{yaron} The distance dependence of the charge-charge screening
arising from the dielectric model of Eq.~(\ref{Eq:7}) is similar to
that obtained from the explicit solvent model of Moore and
Yaron\cite{yaron}, suggesting that the longitudinal dipoles provide a
reasonable description of the dielectric medium.

\begin{table}[tb]
\caption{ Unscreened energies, the transition moment, $\langle
\hat{d}\rangle$, and
the dipole `polarizability', $\alpha = \frac{\langle
\hat{d}\rangle^2}{E(1^1B_u^-)}$ of a $50$-site polymer chain modelled by the P-P-P model. The energies are in eVs. }
\begin{center}
\begin{tabular}{ccccccccc}
\hline \hline $t$ & $U$ & $\delta$ & E($1^1B_u^-$)   & $1^1B_u^-$
binding energy & E($2^1A_g^+$) & E($1^3B_u^+$)  &   $\langle
\hat{d}\rangle$ \AA
 & $\alpha$ \\
\hline
$2.4045$ & $11.13$ & $0.0734$ & $2.85$  & $2.29$ & $2.31$ & $1.28$ & $8.72$ & $26.7$ \\
\hline
\hline
\end{tabular}
\end{center}
\label{Ta:1}
\end{table}

\begin{table}[tb]
\caption{ Screened energies, E, screening energies, S, and the
screened  $1^1B_u^-$ binding energy, BE, of a $50$-site polymer chain. The charge gap is denoted by $2\Delta$.
The energies are in eVs.}
\begin{center}
\begin{tabular}{ccccccccccccc}
\hline \hline $t$ & $U$ & $\delta$ & $\lambda$ & E($1^1B_u^-$) &
S($1^1B_u^-$) & E($2^1A_g^+$) & S($2^1A_g^+$)  & E($1^3B_u^+$) &
S($1^3B_u^+$) & $2\Delta$ & S($2\Delta$) & BE
\\
\hline
$2.4045$ & $11.13$ & $0.0734$ & $0.14$ &
$2.16$  & $0.69$ & $2.22$ & $0.09$ & $1.25$ & $0.03$ & $3.53$ & $1.62$ & $1.37$ \\
\hline \hline
\end{tabular}
\end{center}
\label{Ta:2}
\end{table}

The excitation  energy of the $1^1B_u^-$ exciton is reduced by
$0.69$ eV, which is roughly twice as large as the Yaron-Moore
result\cite{yaron}, and much larger than the DFT-GWA-BSE
calculation, which predicts a negligible screening of the
exciton\cite{bobbert}. The binding energy is therefore reduced
from $2.3$ eV to $1.4$ eV. As already noted, for these parameter
ranges the  $1^3B_u^+$ state is a magnon and the $2^1A_g^+$ state
is a bi-magnon. Both states are therefore strongly covalent with
very little ionicity and thus there is very little screening. This
affect can be traced back to  the monopole-dipole interaction,
Eq.\ (\ref{Eq 7}), which shows that the electric dipole couples to
charge density (or ionic) fluctuations in the wavefunction. These
results are summarized in Table II.

\section{Results}

Having parameterized the model, we now investigate other parameter
ranges. $\lambda$ is found using Eq.\ (\ref{eq:8}) for fixed $\rho$
and $r_1$ with the recalculated value of $\langle \hat{d}\rangle$.
First, we consider $U=10$ eV,  $t=2.5$ eV and $\delta=0.1$, as
this corresponds to the optimal parameter set used by Barford and
Bursill in their study of trans-polyacetylene\cite{barfordPA}. The
single chain P-P-P excitation energies are shown in Table III.
Notice that, unlike the case for the original P-P-P model
parametrization, the $2^1A_g^+$ state lies higher in energy than
the $1^1B_u^-$ state. The screened energies are shown in Table IV,
where we see that the charge gap is solvated by ca.\ $2.0$ eV, and
the $1^1B_u^-$ and $2^1A_g^+$ states are screened by $0.80$ eV and
$0.75$ eV, respectively.  The binding energy of the $1^1B_u^-$
exciton is reduced from $2.3$ to $1.1$ eV.

We next investigate the r\^ole played by the strength of the
Coulomb interactions, particularly on the screening of the
$2^1A_g^+$ state, which evolves from a bi-magnon to the $n=2$
Mott-Wannier singlet exciton, and the triplet, $1^3B_u^+$, state,
which evolves from a magnon to the $n=1$ Mott-Wannier triplet
exciton, as $U$ is reduced\cite{barford02a, barford02b}. We take a
value for $\delta$ of $0.2$ as this maps the linear chain band
structure onto the band structure of poly(p-phenylene). Table IV
confirms the expected trend that as the ionicity of the $2^1A_g^+$
and $1^3B_u^+$ states increases relative to their covalency their
screening, relative to the ionic $1^1B_u^-$ state, increases.

\begin{table}[tb]
\caption{ Unscreened energies, the transition moment, $\langle
\hat{d}\rangle$, and
the dipole `polarizability', $\alpha = \frac{\langle
\hat{d}\rangle^2}{E(1^1B_u^-)}$ of a $50$-site polymer chain modelled by the P-P-P model. The energies are in eVs. }
\begin{center}
\begin{tabular}{ccccccccc}
\hline \hline $t$ & $U$ & $\delta$ &E($1^1B_u^-$)   & $1^1B_u^-$
binding energy  & E($2^1A_g^+$) & E($1^3B_u^+$)  &    $\langle
\hat{d}\rangle$ \AA & $\alpha$
\\
\hline
$2.5$ & $10$ & $0.1$ & $2.67$  & $2.32$ & $2.95$ & $1.69$ & $9.41$ & $33.2$ \\
$2.5$ & $10$ & $0.2$ & $3.69$  & $3.05$ & $4.69$ & $2.76$ & $8.15$ & $18.0$\\
$2.5$ & $8$ & $0.2$ & $3.20$  & $2.76$ & $4.42$ & $2.82$ & $8.81$ & $24.3$  \\
$2.5$ & $6$ & $0.2$ & $2.93$  & $2.42$ & $4.10$ & $2.79$ & $9.14$ & $28.5$ \\
\hline
\hline
\end{tabular}
\end{center}
\label{Ta:3}
\end{table}

\begin{table}[tb]
\caption{ Screened energies, E, screening energies, S, and the
screened  $1^1B_u^-$ binding energy, BE, of a $50$-site polymer chain. The charge gap is denoted by $2\Delta$.
The energies are in eVs.}
\begin{center}
\begin{tabular}{ccccccccccccc}
\hline \hline $t$ & $U$ & $\delta$ & $\lambda$ & E($1^1B_u^-$)   &
S($1^1B_u^-$)  & E($2^1A_g^+$) & S($2^1A_g^+$)  &    E($1^3B_u^+$)
& S($1^3B_u^+$) & $2\Delta$ & S($2\Delta$) & BE
\\
\hline
$2.5$ & $10$ & $0.1$ & $0.151$ & $1.87$  & $0.80$ & $2.20$ & $0.75$ & $1.59$ & $0.10$ & $3.00$ & $1.99$ & $1.13$ \\
$2.5$ & $10$ & $0.2$ & $0.131$ & $3.22$  & $0.47$ & $4.34$ & $0.35$ & $2.73$ & $0.03$ & $5.43$ & $1.31$ & $2.21$ \\
$2.5$ & $8$ & $0.2$ & $ 0.141$ & $2.60$  & $0.60$ & $3.45$ & $0.97$ & $2.71$ & $0.11$ & $4.26$ & $1.70$ & $1.66$ \\
$2.5$ & $6$ & $0.2$ & $0.147$ & $2.19$  & $0.74$ & $2.48$ & $1.62$ & $2.40$ & $0.39$ & $3.26$ & $2.09$ & $1.07$ \\
\hline
\hline
\end{tabular}
\end{center}
\label{Ta:4}
\end{table}

Length scales also play a r\^ole in the screening of the excited
states and the charge gap. As the separation between a polymer and
a line of dipoles, $r$, is increased the typical distance in the
monopole-dipole interaction increases.  When this interaction
distance is larger than the particle-hole separation, the charge
character of the exciton approaches that of a point dipole. For
the line dipole used here, the ratio of the screening energy of
the $1^1B_u^-$ state versus the charge gap is $0.6$ when $r = 4$
$\AA$ and reduces to a roughly constant value of $0.4$ at $r \geq
16$ $\AA$.  This indicates that a substantial fraction of the
screening energy of the $1^1B_u^-$ exciton is from dipole-dipole
interactions.

\section{Conclusions}

In this paper we introduced a dynamical model of a dielectric to
investigate the screening of intra-chain excitations by the
environment. This model consisted of a linear array of quantized
longitudinal dipoles interacting with a conjugated polymer,
represented by the Pariser-Parr-Pople model. The dipole parameters
are adjusted to provide a realistic model of the dielectric. The
model reproduces the expected solvation energy of a point charge
in these materials and sets the frequency of the dielectric
response to the optical gap of the material. The model is
one-dimensional and is solved by the density matrix
renormalization group method. The one-dimensionality of the model
does mean that it is impossible to include dipole-dipole
interactions, and thus the model misses some of the physical
characteristics of real dielectric materials.

We make the following conclusions from our results.  (1) The
$1^1B_u^-$ exciton binding energy is typically reduced by ca.\ 1
eV to just over 1 eV.  Thus, this relatively simple
one-dimensional model of the screening of conjugated polymers by
the environment predicts $1^1B_u^-$ binding energies rather close
to the experimentally accepted solid state values of ca.\ $1$ eV.
(2) Covalent (magnon and bi-magnon) states are very weakly
screened compared to ionic (exciton) states. (3) Screening of the
$1^1B_u^-$ exciton is closer to the dispersion than solvation
limit.

\begin{acknowledgements}
We thank Eric E.\ Moore for useful discussions.  This work was
supported by the U.\ K.\ Engineering and Physical Sciences
Research Council (GR/R03921). R.\ J.\ B.\ is supported by the
Australian Research Council and the J G Russell Foundation. D.\
Y.\ J.\ is supported by the US National Science Foundation
(0316759).
\end{acknowledgements}


\begin{references}
\bibitem{footnote3} Or more correctly, the time-scale for the electron-hole motion is inversely proportional to
 the energy gap to the next
dipole-connected excited state (e.\ g.\ $(E(mA_g) - E(1B_u))$ for
centro-symmetric polymers).
\bibitem{footnote6} In principal, the model could be extended to include transverse dipoles by including an
additional flavor of boson in the model described in Eq.
(\ref{eq3}).
\bibitem{footnote4}  The transition moment operator is defined as ${\hat d} = \sum_i ({\hat N_i}-1)x_i$.  Thus,
 the transition moment associated with the $1^1B_u^-$ state, $\langle {\hat d} \rangle$,
 is $\langle 1^1B_u^-|{\hat d}| 1^1A_g^+ \rangle$.
\bibitem{footnote1} One consequence of this symmetry is that the electronic ground state is associated with a
manifold of `vibrational' states, with a spacing $\hbar\omega_0$
and  alternating between $^1A_g^+$ and  $^1B_u^-$ symmetry. Thus,
the $1^1B_u^-$ state does not necessarily correspond to the lowest
`vibrational' state of the excited exciton. We check that it does
by calculating its dipole moment to the $1^1A_g^+$  state.
\end{references}
\end{document}